\documentclass[aps,prl,twocolumn,showpacs,preprintnumbers,amsmath,amssymb,nofootinbib]{revtex4-1}
\usepackage{tikz}

 \def\ep{{\epsilon}}

 \def\frac#1#2{{#1\over #2}}

 \def\G{{\Gamma}}

\def\be{\begin{equation}}
\def\ee{\end{equation}}
\def\bea{\begin{eqnarray}}
\def\eea{\end{eqnarray}}

\newcommand{\Op}{\mathcal{O}}

\def\linka{\tikz[baseline=.0ex,scale=0.8]{\draw[thick] (0,1/2) arc (-300:40:1/2);\draw[->,thick] (0,1/2) arc (-300:25:1/2);\draw[thick] (-0.05,-0.2) arc (210:-122:1/2);\draw[->,thick] (-0.05,-0.2) arc (210:152:1/2);}}

\def\traza{\tikz[baseline=.0ex,scale=0.7]{\draw[->,thick] (0,1/2) arc (-360:30:1/2);)}}

\def\trazai{\tikz[baseline=.1ex,scale=0.7]{\draw[<-,thick] (0,0.75) arc (-210:150:1/2);)}}

\def\nn{\nonumber \\ }

 \def\ep{\epsilon}

\usepackage{color}
\usepackage{graphicx}
\usepackage{dcolumn}
\usepackage{bm}

\def\f{\frac}
\def\ba{\begin{eqnarray}}
\def\ea{\end{eqnarray}}
\newcommand{\braket}[1]{\left<#1\right>}
\def\mcl{\mathcal}

\begin{document}

\begin{flushright}                                      
NORDITA-2016-6
YITP-16-10
\end{flushright} 
\title{Scrambling without chaos in RCFT}

\author{Pawe{\l} Caputa$^{a}$, Tokiro Numasawa$^b$ and Alvaro Veliz-Osorio$^{c,d}$}

\affiliation{$^a$Nordita and Stockholm University, Roslagstullsbacken 23, SE-106 91 Stockholm, Sweden\\
$^b$Yukawa Institute for Theoretical Physics (YITP), Kyoto University, Kyoto 606-8502, Japan\\
$^{c}$ Mandelstam Institute for Theoretical Physics, School of Physics University of the Witwatersrand, Johannesburg, WITS 2050, South Africa\\
$^{d}$ School of Physics and Astronomy, Queen Mary, University of London, Mile End Road, London E1 4NS, United Kingdom}


\begin{abstract}
In this paper we investigate measures of chaos and entanglement in rational conformal field theories in 1 + 1 dimensions. First, we derive a formula for the late time value of the out-of-time-ordered correlators for this class of theories. Our universal result can be expressed as a particular combination of the modular S-matrix elements known as anyon monodromy scalar. Next, in the explicit setup of an $SU(N)_k$ Wess–Zumino–Witten model, we compare the late time behavior of the out-of-time-ordered correlators and the purity. Interestingly, in the large-$c$ limit, the purity grows logarithmically as in holographic theories; in contrast, the out-of-time-ordered correlators remain, in general, nonvanishing.

\end{abstract}

\maketitle

\section{Introduction}
Two-dimensional conformal field theories (2d CFTs) have played an important role in understanding a number of interesting 
questions in theoretical physics. In this vein they've become central tools in the study of  entanglement \cite{Calabrese:2004eu} and more recently quantum chaos. Based on earlier work on superconductors by Larkin and Ovchinnikov \cite{LO}, Kitaev has proposed that chaotic behavior in quantum systems can be diagnosed by computing the expectation value of  the square of commutators of local operators \cite{Kitaev} . This essentially amounts to calculating the out-of-time order thermal correlator (OTOC)
\be\label{OTO}
C_{ij}^{\beta}(t) \equiv \f{\braket{\mcl{O}_i^{\dagger}(t)\mcl{O}^{\dagger}_j\mcl{O}_i(t)\mcl{O}_j}_\beta}{\braket{\mcl{O}_i^{\dagger}\mcl{O}_i}_\beta \braket{\mcl{O}_j^{\dagger}\mcl{O}_j}_\beta}\,.
\ee
If this quantity vanishes exponentially at late times for generic operators then the quantum system is chaotic. A number of universal properties of this object can be obtained for 2d CFTs. In particular, its been argued that chaotic behavior might be a telling characteristic of holographic CFTs \cite{Roberts:2014isa,Roberts:2014ifa,Maldacena:2015waa,Fitzpatrick:2016thx}.

On the other hand, one of the characteristic features of CFTs at large central charge is a so-called scrambling of entanglement  \cite{Asplund:2015eha}. One particular incarnation of scrambling is a logarithmic evolution of R\'enyi entanglement entropies after local operator excitation. Here we will focus on the second R\'enyi entropy or simply the purity. Various studies showed that, for rational CFTs (RCFTs), the purity saturates to a constant equal to the logarithm of the quantum dimension of the local operator's conformal family \cite{Nozaki:2014hna,Caputa:2015tua, Caputa:2016yzn}. Meanwhile, it is believed that in holographic CFTs (consistent with Ryu-Takayanagi formula \cite{Ryu:2006bv}) the R\'enyi entropies will grow logarithmically with time  \cite{Caputa:2014vaa, Asplund:2015eha} (also at large-c, the scrambling time can be naturally obtained in a similar setup from the evolution of the mutual information in CFT and holographically \cite{Shenker:2013pqa,Caputa:2014eta,Caputa:2015waa}\footnote{The relation between scrambling and OTO correlators was also demonstrated in \cite{Hosur:2015ylk} but the connection with entanglement scrambling \cite{Asplund:2015eha} used here is not clear to us at this moment}.). This means that in large c, holographic CFTs, the information about non-perturbative constants (like quantum dimensions or modular S-matrix)  gets scrambled.

In this work we would like probe the similarities and differences between purity and OTOCs in the setup of RCFTs and find out which specific (non-perturbative) information about the theory is forfeit by quantum chaos. For that, we first fill the existing gap and compute the late time value of the OTOCs valid for any RCFT. Next, we consider a non-trivial integrable 2d CFT, the $SU(N)_k$ Wess-Zumino-Witten (WZW) model, where a number of known results can be put in the new light of  entanglement and quantum chaos measures. Moreover, we consider a large-$c$ 't Hooft limit that shares some features with holographic CFTs and compare the evolution of purity and OTOC in this regime. We observe that, in the large-$c$ limit, the purity grows logarithmically, while the OTOCs approach a non-vanishing constant value.

This letter is organized as follows: In Sec. 2, we compute the late time value of OTOC in RCFT and topological quantum field theory (TQFT).  In Sec. 3, we revise the relationship between purity, quantum dimension and logarithmic growth. In Sec. 4, illustrate both OTOC and purity for a $SU(N)_k$ WZW model. Finally, in Sec. 5 study the behavior of these quantities in the 't Hooft limit. Finally, we conclude and place details in two appendices.

\section{ Late time of OTOC in RCFTs}

In the present section we compute the late time value of the OTOC \eqref{OTO} with insertion points \cite{Roberts:2014isa}
\bea
z_1&=&e^{\frac{2\pi}{\beta}(t+i\epsilon_1)},\qquad \bar{z}_1=e^{-\frac{2\pi}{\beta}(t+i\epsilon_1)},\nn
z_2&=&e^{\frac{2\pi}{\beta}(t+i\epsilon_2)},\qquad \bar{z}_2=e^{-\frac{2\pi}{\beta}(t+i\epsilon_2)},\nn
z_3&=&e^{\frac{2\pi}{\beta}(x+i\epsilon_3)},\qquad \bar{z}_3=e^{\frac{2\pi}{\beta}(x-i\epsilon_3)}, \nn
z_4&=&e^{\frac{2\pi}{\beta}(x+i\epsilon_4)},\qquad \bar{z}_4=e^{\frac{2\pi}{\beta}(x-i\epsilon_4)}.
\eea 
The main message from these points is that for the appropriate ordering of epsilons $\epsilon_i$ (see the figures) as we increase $t$ the cross-ratio $z=(z_{12}z_{34})/(z_{13}z_{24})$ encircles clockwise the point $z=1$ in the complex plane and comes back to $0$ (this doesn't happen with $\bar z$). The role of the temperature in this specific behavior of $z$ is not crucial and it  is only used to extract the universal predictions for the quantum chaos. More precisely, in chaotic CFTs, these correlators are expected to damp after the so-called scrambling time \cite{Roberts:2014ifa}. In contrast, for RCFTs, which are integrable systems, one expects $C^\beta_{ij}(t)$ to reach constant non-vanishing values. Indeed, as we shall see, the OTOCs are given by the succinct formula
\be
C^\beta_{ij}(t)\to \frac{1}{d_id_j}\frac{S^*_{ij}}{S_{00}}\label{OTOLT}\, 
\ee
at late times, where $S^*_{ij}$ is the complex conjugate of the modular S-matrix. 
The argument proceeds as follows, first we write 
\ba
\braket{\mcl{O}^{\dagger}_i(z_1,\bar{z}_1)\mcl{O}_i(z_2,\bar{z}_2)\mcl{O}^{\dagger}_j(z_3,\bar{z}_3)\mcl{O}_j(z_4,\bar{z}_4)} \notag \\ 
= |z_{12}|^{- 4h_i} |z_{34}|^{-4 h_j} f(z,\bar{z}).
\ea
Then, we express $f(z,\bar{z})$ in terms of the conformal blocks of the theory $\mcl{F}^{ii}_{jj}(p|z)$ (and their anti-holomorphic counterparts $\bar{\mcl{F}}^{ii}_{jj}(p|\bar{z})$)
\be\label{conformal block}
f(z,\bar{z}) = \sum_p \mcl{F}^{ii}_{jj}(p|z)\bar{\mcl{F}}^{ii}_{jj}(p|\bar{z}).
\ee

At early times, since $z \approx 0 $ and $\bar{z} \approx 0$, the contribution from the identity channel ($p=0$) dominates; thus, $f(z,\bar{z})\approx 1$.
At late times, once again $z \approx 0 $ and $\bar{z} \approx 0$. However, as time goes by, the cross-ratio $z$ traverses a non-trivial contour around $z=1$ in the complex plane (this is not the case for $\bar z$). As shown in \cite{Roberts:2014ifa}, extracting this monodromy from the explicit form of the large-$c$ conformal block \cite{Fitzpatrick:2014vua} one can see the butterfly effect in 2d CFT. In RCFTs the monodromy of conformal blocks is given by a finite matrix and we have
\be
\mcl{F}^{ii}_{jj}(p|z) \to \sum_q \mcl{M}_{pq} \mcl{F}^{ii}_{jj}(q|z)\,.
\ee
Because cross ratio $z$ goes around $z=1$ and finally comes back to $z=0$, the only relevant component  is $\mcl{M}_{00}$. Therefore, we obtain 
\be
\lim_{t \to \infty} f(z,\bar{z}) = \mcl{M}_{00} \mcl{F}^{ii}_{jj}(0|z) \bar{\mcl{F}}^{ii}_{jj}(0|\bar{z}).
\ee
Moreover, for RCFTs this monodromy matrix element can be expressed in terms of the modular $S$-matrix as \cite{Moore:1988ss}:
\be
\mcl{M}_{00}= \f{S^*_{ij}}{S_{00}} \f{S_{00}}{S_{0i}}\f{S_{00}}{S_{0j}}\,. \label{Minimalmonodromy}
\ee
\begin{figure}
\begin{center}
\includegraphics[scale=0.29]{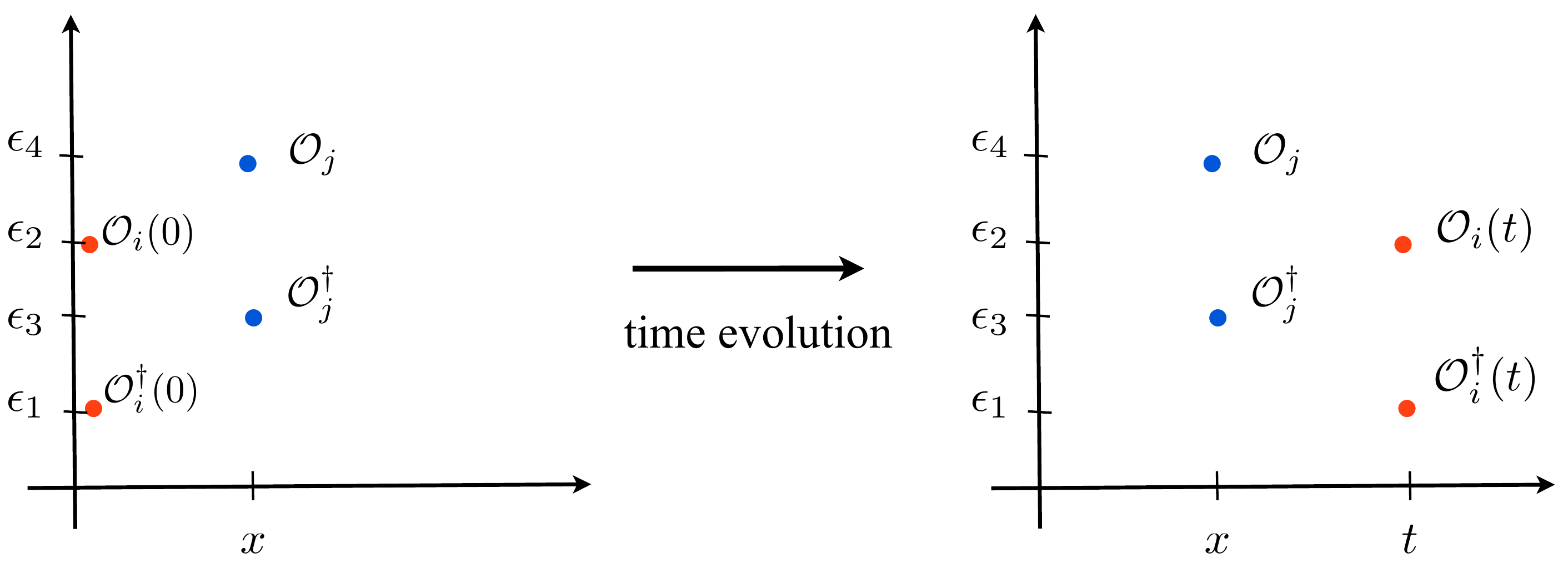}
\end{center}
\caption{The orbit of chiral part of operators in complex plane}
\label{timee}
\end{figure}

\begin{figure}
\begin{center}
\includegraphics[scale=0.27]{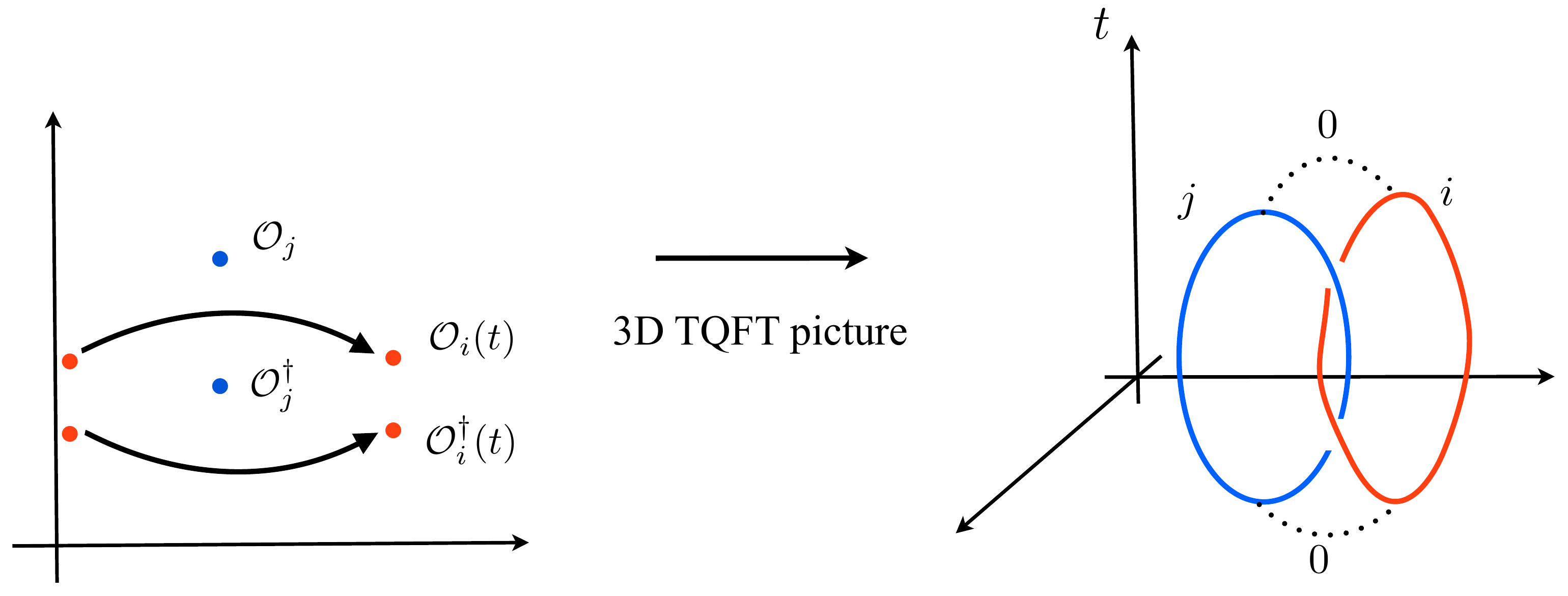}
\end{center}
\caption{3d TQFT counterpart. Here, $0$ denotes the identity channel of the conformal blocks.
}
\label{tqft}
\end{figure}

We can also derive this late time value of the OTOC using $3d$ TQFT technology \cite{Witten:1988hf}. 
As time passes, the operators evolve as depicted in Fig.\ref{timee}.
Their orbits are mapped to 3d links made by the corresponding anyons as in Fig.\ref{tqft}. 
The relation between 2d CFT and 3d TQFT is given as follows. 
First, the initial state of 3d TQFT is determined by the sector of conformal block we choose. 
In this case we choose the identity sector in CFT and in 3d TQFT the pairs of anyons are created from the vacuum. Then, because there is a monodromy in CFT side, there is a link in 3d TQFT side. 
Finally, corresponding to taking the identity sector at late time, anyons fuse to the vacuum, which means that the final state in the 3d TQFT is given by the pair annihilation of anyons. 
As a result, we obtain the Hopf link of two Wilson loops. 
From this observation, we find that the monodromy matrix element is given by the expectation value of the Hopf link divided by the expectation value of two non-linked Wilson loops.
Based in results from \cite{Witten:1988hf}, we find
\be\label{OTO late}
C^{\beta}_{ij}(t)\to\frac{{}_i\linka{}_j }{{}_i \;\traza \;\trazai\; {}_j}\equiv\frac{1}{d_id_j}\frac{S^*_{ij}}{S_{00}}.
\ee
This exactly matches with the r.h.s of (\ref{Minimalmonodromy}) and naturally explains why this combination appears in late time OTOC. If we apply this formula to the Ising model CFT, we reproduce exactly the results from the explicit calculation of monodromy in Appendix B of \cite{Roberts:2014ifa}.

Let us finally mention that the above late time value, known as monodromy scalar, has been proposed as a measure of non-abelian anyons in interferometry experiments \cite{Bonderson}. It would be interesting to explore this connection as a possible experimental measure of quantum chaos.

\section{Purity and quantum dimension}

Now, we turn our attention to entanglement. We are interested in a local quench setup where a state is excited by a local operator. More precisely, we take a pure state
 in a 1+1 dimensional CFT and divide space into two halves $A$ and $\bar A$. 
Then, we insert a local operator ${\cal O}$ with conformal dimension $h=\bar{h}$ into $\bar A$ at, say $x=-l$, and study the time evolution of entanglement in the system. In particular, we consider the evolution of the second R\'enyi entropy; hereafter we refer to this quantity as the purity (strictly speaking the purity corresponds to the logarithm of the square of the reduced density matrix). Using the replica method the purity can be extracted from the canonical 4-point function $\mathcal{G}(z,\bar{z})\equiv\langle \Op(0)\Op(z,\bar{z})^\dagger\Op(1)\Op^\dagger(\infty)\rangle$ and it reads \cite{He:2014mwa} 
\be\label{second renyi}
\Delta S^{(2)}_A(z,\bar{z})=-\log\left[ |z(1-z)|^{4h}{\cal G}(z,\bar{z})\right] \,,
\ee 
where the points entering the cross-ratios are expressed in terms of the replica points as $z^2_i=w_i$ where
\bea
&& w_1=i(\epsilon -it)-l, \ \ w_2 = -i(\epsilon+it)-l, \nn
&& \bar{w}_1=-i(\ep-it)-l,\ \ \bar{w}_2=i(\epsilon+it)-l.\label{points}
\eea
As one takes $\epsilon\to 0$, $\bar z\to 0$, meanwhile, $z$ can become either $0$ or $1$ for times earlier or later than $l$ respectively. 

 In a RCFT given the singularity structure of ${\cal G}$ this implies that $\Delta S^{(2)}_A$ vanishes at early times since only the identity channel contributes. Moreover, since early and late times are mapped to each other by the transformation $(z,\bar z)\to (1-z,\bar z)$ one finds that the late time purity can be extracted from the fusion matrix element $F_{00}[\Op]$. Furthermore, this quantity corresponds to the inverse of the quantum dimension of ${\cal O}$'s conformal family. Hence, at late times we have \cite{He:2014mwa,Caputa:2015tua}
\be
\Delta S^{(2)}_A(t)=\log d_{\cal O}\,.
\ee
Observe that the appearance of a constant contribution at late time for $\Delta S^{(2)}_A$ is closely related to the singular behavior 
\be
{\cal G}(z,\bar{z})\to  d^{-1}_\Op((1-z)\bar{z})^{-2h}
\ee
 of the four-point function as $(z,\bar{z})\to(1,0)$. The authors of \cite{Asplund:2015eha} have argued that in holographic CFTs, where 
the Ryu-Takayanagi formula \cite{Ryu:2006bv} is valid, such singularity disappears due to ``scrambling of entanglement". This way, in our setup, the appearance of a the (non-perturbative) quantum dimension at late times is replaced by a divergent logarithmic growth of the R\'enyi entropy. This fast growth of entanglement is equivalent to the breakdown of the quasi-particle picture that is characteristic of strongly coupled large-c theories. 
The temperature dependence can be introduced by the standard conformal map (see \cite{Caputa:2014eta}) but the logarithmic growth with time at large c is not affected. Below, we will show how this behavior emerges in the ``holographic"  large-$c$ limit of WZW models.  

It is also worth mentioning that the $\log d_\Op$ increase can be obtained from the topological entanglement entropy \cite{wen2} if one of the regions contains an anyonic excitation \cite{Dong:2008ft}. At first sight the late time purity and the OTOC are rather similar objects i.e. both are captured by the vacuum conformal block. Naively, one would expect to be able to use them interchangeably as indicators of quantum chaos, or even to diagnose whether a CFT has a holographic dual. This is in fact not the case and below we
present an example where the purity displays the behavior expected from a holographic theory, while the OTOC doesn't.

\section{On purity and OTOC in $SU(N)_k$ WZW}

In this section we consider a WZW model with affine Lie algebra $SU(N)_k$. Just as the quantum dimension, the late time value OTOC is invariant under level-rank duality, hence, the following discussion is valid for  $SU(k)_N$ as well. Knowing the modular S-matrix of the model, using \eqref{OTO late} we could compute directly the late time OTOC. Here we follow a direct approach instead, both to illustrate the underlying mechanisms and as a consistency check. We focus on the 4-point function of operators $g^{\alpha}_\beta(z_i,\bar{z}_{i})$ (and their conjugates)  in the fundamental representation $\alpha=\{1,0,...0\}$ that have conformal dimension
\be
 h=\bar{h}=\frac{N^2-1}{2N \kappa}.
 \ee
 where $\kappa=N+k$. The general correlator that we employ is
 \bea\label{g4}
 &&\langle g^{\alpha_1}_{\beta_1}(z_1,\bar{z}_1)(g^{-1})^{\beta_2}_{\alpha_2}(z_2,\bar{z}_2)g^{\alpha_3}_{\beta_3}(z_3,\bar{z}_3)(g^{-1})^{\beta_4}_{\alpha_4}(z_4,\bar{z}_4)\rangle\nn
 &&\equiv\frac{1}{|z_{12}|^{4h}|z_{34}|^{4h}}|z|^{4h}\mathcal{G}(z,\bar{z}).
 \eea
Recall that we characterized OTOCs by the function $f(z,\bar{z})$ which is related to $\mathcal{G}(z,\bar{z})$ via
$f(z,\bar{z})= |z|^{2h}\mathcal{G}(z,\bar{z})$. 
To apply the above correlator in our OTOC we set $\alpha_1=\alpha_2$, $\beta_1=\beta_2$ and $\alpha_3=\alpha_4$ with $\beta_3=\beta_4$. On the other hand, for the purity all the $\alpha$s (and $\beta$s) and  must be equal.

 The general 4-point functions \eqref{g4} are well known solutions of the Knizhnik-Zamolodchikov equations \cite{Knizhnik:1984nr}.
 The canonical correlator can be expanded in terms of affine conformal blocks
 \be\label{G}
 \mathcal{G}(z,\bar{z})=\sum_{i,j}I_i\bar{I}_j\sum_{n}X_{nn}\mathcal{F}^{(n)}_{i}(z)\mathcal{F}^{(n)}_j(\bar{z}),
 \ee
 with $i,j,n\in\{1,2\}$ and $SU(N)$ factors $I_1=\delta^{\alpha_2}_{\alpha_1}\delta^{\alpha_4}_{\alpha_3}$, $I_2=\delta^{\alpha_4}_{\alpha_1}\delta^{\alpha_2}_{\alpha_3}$. In our arguments we will only use $X_{11}=1$, more details can be found in \cite{DiF}.

Let us compute the purity first. In order to extract the late time value, we apply the fusion transformation that mixes conformal blocks 
\be\label{crossing holomorphic}
{\cal G}(1-z,\bar{z})=\sum_{i,j}I_i\bar{I}_j\sum_{n,m}X_{nn}c_{nm}\mathcal{F}^{(m)}_{3-i}(z)\mathcal{F}^{(n)}_j(\bar{z})\,,
\ee
where the relevant coefficient is
\be\label{c plus plus}
c_{11}=N\frac{\G(N/\kappa)\G(-N/\kappa)}{\G(1/\kappa)\G(-1/\kappa)}=[N]^{-1}=d^{-1}_g,
\ee
with $ d_g$ being the quantum dimension for the fundamental representation, where the quantum numbers are defined as
\be
[x]=\frac{q^{x/2}-q^{-x/2}}{q^{1/2}-q^{-1/2}}\qquad q=e^{-\frac{2\pi i}{N+k}}\, .
\ee
Taking the limit of the conformal blocks for $(z,\bar{z})\to(1,0)$ (see App.A) leaves us with the log of the quantum dimension multiplied by the appropriate singularity such that we get the $\log\, [N]$ at late times. It is also interesting to see that even though the four-point correlator is expanded in terms of the affine conformal blocks, that are sums of the Virasoro blocks, the relevant constant is still hidden in the vacuum block. Moreover, from the definition, we have $[N]=[k]$ which is in fact the consequence of the level-rank duality for quantum dimensions inherited by the purity.


Now, let us study the OTOC. Extracting the monodromy around $z=1$ brings us to
\be
f(z,\bar{z})=e^{-2\pi i(h_\theta-2h)}\sum_{i,j}I_i\bar{I}_j\sum_{n,m}X_{nn}B_{nm}f^{(m)}_{i}(z)f^{(n)}_j(\bar{z}).
\ee
where $B_{nm}$ are the monodromy matrix elements of the solutions of the hypergeometric equation (see e.g. \cite{Erdelyi}). Taking the limit of $(z,\bar{z})\to(0,0)$  leaves only the terms from $f^{(1)}_1$ and we are left with the overall exponent prefactor and the coefficient $B_{11}$ given by
\bea
B_{11}=1-2ie^{-i\pi(1-\frac{N}{\kappa})}\frac{\sin^2(\frac{\pi}{\kappa})}{\sin(\pi(1-\frac{N}{\kappa}))}.
\eea
After some algebra, and expressing the answer in terms of quantum numbers we find that at late times
\be
C^\beta_{ij}(t)\to e^{-2\pi i (h_\theta-2h)}B_{11}=q^{\frac{1}{N}+\frac{1}{2}}\frac{\left(q^{-\frac{N+2}{2}}+[N-1]\right)}{[N]}.\label{LTV}
\ee
We can compare this answer with our RCFT result Eq.\,\eqref{OTOLT}. Indeed, the S-matrix element for the present example has been computed in \cite{Dong:2008ft} and it reads
\be
\frac{S^*_{\alpha\alpha}}{S_{00}}=q^{\frac{1}{N}+\frac{1}{2}}\left(q^{-\frac{N+2}{2}}+[N-1]\right)[N]\, ,
\ee
and inserting $d_i=d_j=[N]$ beautifully matches \eqref{OTOLT}. For example, for the $SU(2)_k$ model, the late time OTO \eqref{LTV} reduces to 
\be
C^\beta_{ij}(t)\to\cos\left(\frac{2\pi}{k+2}\right)\cos^{-1}\left(\frac{\pi}{k+2}\right)\,,
\ee
 which can be extracted from the explicit form of the $SU(2)_k$ modular S-matrix
\be
S_{ij}=\sqrt{\frac{2}{2+k}}\sin\left(\frac{(i+1)(j+1)\pi}{k+2}\right)\,,
\ee 
by setting $i=j=1$.
Note that, in general, the elements of the modular S-matrix can be complex (except the first row that are related to quantum dimensions which are real).

Summarizing, we have shown that late time values of the purity and OTOC are given in terms of the quantum dimensions as well as the modular S-matrix. It is interesting that, in RCFTs, OTOCs give us the access to the entire modular S-matrix whereas R\'enyi entropies only to the first row $S_{0i}$. It is also interesting to consider the classical limit ($k\to\infty$) of WZW models where the purity becomes the log of the dimension of the fundamental, and the OTOC equals one. 

\section{OTOC and purity in the large-c limit}

Finally, it is interesting to compare the behavior of the purity and the OTOC in the large-$c$ limit. In the $SU(N)_k$ WZW the central charge is given by 
\be
c=\frac{k(N^2-1)}{k+N}\,.
\ee
 By introducing the 
 't Hooft coupling constant
\be\label{'t Hooft coupling}
\lambda=\frac{N}{k}\,,
\ee
we can define a 't Hooft limit of large central charge with the coupling fixed (weak or strong). The four-point correlator has been analyzed in detail in this limit by \cite{Kiritsis:2010xc} and we apply their analysis in our context. For $c\to\infty$, the 4-point correlator becomes (see App A) (Note that here, unlike in \cite{Roberts:2014ifa}, all our operators are light: $h/c\to 0$ as $c\to \infty$). Using this correlator, one can see that for a large central charge the singularities leading to the quantum dimension are absent, which leads to a logarithmic growth of the purity
\be\label{c infinite}
\Delta S^{(2)}_A(t)\simeq 
2h\log\left(\frac{2t}{\epsilon}\right)-\log(2).
\ee
This behavior comes from discarding terms proportional to $\frac{1}{\sqrt{c}}$.
However, if we include such corrections, then the late time answer becomes the logarithm of the quantum dimension in the large-$c$ limit. 
It is illustrative to verify this in the strong coupling regime, where $h=1/2$ and the correlator (see App A) can be computed by approximating the operators $g^{\alpha}_{\beta}(z_i,\bar{z}_i)\simeq\frac{1}{k}\sum^k_{i=1}\psi^\alpha(z_i)\bar{\psi}^\beta(\bar{z}_i)$ with complex fermions. In this limit we have
\be
\mathcal{G}(z,\bar{z})\simeq\frac{I_1\bar{I}_1}{|z|^2}+\frac{I_2\bar{I}_2}{|1-z|^2}+\sqrt{\frac{\lambda}{c}}\left(\frac{I_1\bar{I}_2}{z(1-\bar{z})}+\frac{I_2\bar{I}_1}{(1-z)\bar{z}}\right).\label{GLc}
\ee
Using \eqref{second renyi}, it is clear that neglecting the last two terms in the above expression leads to the logarithmic growth of the purity in the large-c limit, which is sometimes known as {\it scrambling of entanglement}. Another way to look at this order of limits issue is that, at strong coupling, the time scale at which the purity reaches the $\log d_O$ can be estimated as $t-l\simeq \frac{c^{1/4}}{2\lambda^{1/4}}\epsilon$. If we then take the large-c limit first (like in holography), we will not reach the finite constant and we are left with the logarithmic growth with time (see also discussion in \cite{Caputa:2014vaa}).

On the other hand, the late time value of the OTOC comes from the first term in \eqref{GLc} (irrespectively of the weak or strong coupling) and in $f(z,\bar{z})$ it is simply 1. For different operators $\alpha_1=\alpha_2\neq\alpha_3=\alpha_4$ only $I_2$ vanishes so the result remains the same. Thus, OTOC is a good indicator of integrability even when entanglement scrambles.

\section{Conclusion}
We have shown that OTOC in RCFTs approach to a universal constant at late times which is completely determined in terms of the modular S-matrix of the theory. Moreover, we have pointed out that this quantity is potentially observable in experimental set-ups. We provided a non-trivial example in the integrable $SU(N)_k$ WZW model. We also argued in this setup, that in the large-$c$ limit the purity displays the logarithmic growth characteristic of holographic models, but the OTOCs remain constant, as a good measure of quantum chaos should since the theory is far from chaotic. It would be interesting to understand how chaos and scrambling are related and emerge in large-c CFT with holographic duals.  Extending our work to operators in higher representations, non-rational CFTs and (non)integrable theories in higher dimensions might shed more light on these issues.

\section*{Acknowledgment}

We would like to thank Tadashi Takayanagi, Shinsei Ryu, Tadakatsu Sakai, Sachin Jain, Howard Schnitzer, Kostya Zarembo and Vishnu Jejjala for discussions on related topics and especially Yingfei Gu, Tadashi Takayanagi, Diptarka Das and Seyed Morteza Hosseini for comments on the draft. PC is supported by the Swedish Research Council (VR) grant 2013-4329. The work of AVO. is based upon research supported in part by the South African Research Chairs Initiative of the Department of Science and Technology and National Research Foundation. TN is supported by JSPS fellowship. Note added: When preparing this letter for submission we become aware of interesting parallel work by Yingfei Gu and Xiao-Liang Qi on OTO correlators in RCFTs \cite{Gu:2016hoy}. We would like to thank Xiao-Liang Qi for sharing their draft before submission.

\appendix
\subsection{Appendix A: 4-Point function in $SU(N)_k$ WZW}
From KZ equations one can derive the canonical four-point function that is written in terms of the cross-ratios and the dimensions $h=\frac{N^2-1}{2N(k+N)}$ and  $h_\theta=\frac{N}{N+k}$. More precisely, the affine conformal blocks are expressed as
\bea
{\cal F}^{(1)}_1(z)&=&z^{-2h}(1-z)^{h_\theta-2h}u_1(z)\nn
{\cal F}^{(2)}_1(z)&=&z^{-2h}(1-z)^{h_\theta-2h}u_2(z)\nn
{\cal F}^{(1)}_2(z)&=&\frac{1}{k}z^{1-2h}(1-z)^{h_\theta-2h}\tilde{u}_1(z)\nn
{\cal F}^{(2)}_2(z)&=&-N z^{1-2h}(1-z)^{h_\theta-2h}\tilde{u}_2(z)\nonumber
\eea
with $u(z)_i$ and $\tilde{u}_i(z)$ being the standard solutions of the hypergeometric equation $u_1(z)=\,_2F_1(a,b,c;z)$ and $u_2(z)=z^{1-c}\,_2F_1(a-c+1,b-c+1,2-c;z),$ where $u_1$ and $u_2$ parametrized by $a=\frac{1}{\kappa}$, $b=-\frac{1}{\kappa}$ and $c=1-\frac{N}{\kappa}$ such that $1-c=\frac{N}{\kappa}=h_\theta$, and $\tilde{u}_1$ and $\tilde{u}_2$ of $a=1+\frac{1}{\kappa}$, $b=1-\frac{1}{\kappa}$ and $c=2-\frac{N}{\kappa}$ such that $1-c=\frac{N}{\kappa}-1=h_\theta-1$. The monodromy of the conformal blocks under the loop that encircles $z=1$ is a combination of the contribution from the pre-factors as well as the monodromy of the hypergeometric functions (see \cite{Erdelyi}). 

The function ${\cal G}(z,\bar z)$ in Eq. \eqref{G} admits the large-$c$ expansion given by
\bea \label{large c G}
{\cal G}(z,\bar z)&\simeq&\frac{ I_1\bar I_1}{|z|^{4h}}
+\frac{I_2\bar I_2 }{|1-z|^{4h}}\\ &+& \frac{\lambda}{\sqrt{c(1+\lambda)}}\Bigg[\Big(\gamma(z,\bar z)\bar I_1 I_2+\text{c.c.}\Big)\Bigg],\nonumber
\eea
where
\be\nonumber
\gamma(z,\bar z)=\frac{\,_2F_1\left(1,1,\frac{2+\lambda}{1+\lambda}; z\right) }{\bar z^{2h} z^{2h-1}}-\frac{\,_2F_1\left(\frac{\lambda}{1+\lambda},\frac{\lambda}{1+\lambda},\frac{1+2\lambda}{1+\lambda};\bar z\right)}{\lambda(1- z)^{2h}} ,
\ee
with $2h=\lambda/(1+\lambda)$.
Notice that around $(z,\bar z)\approx (1,0)$
\be\nonumber
\gamma(z,\bar z)\approx \left(\frac{\pi}{1+\lambda}\right)\frac{\csc\left(\frac{\lambda}{1+\lambda}\right)}{\bar z^{2h}(1-z)^{2h}}\,.
\ee
Plugging the above expression into ${\cal G}$ and afterwards in \eqref{second renyi}, one finds a constant contribution to the late time purity; this constant duly corresponds to the logarithm of the first term in the large-c expansion of the quantum dimension.  

\subsection{Appendix B: Liouville theory}
It is also interesting to ``naively" apply our formula for late time value of OTO in a (non-rational) Liouville 2d CFT with central charge $c=1+6Q^2$. From the explicit form of the analog of the S-matrix \cite{Zamolodchikov:2001ah} (see also \cite{Jackson:2014nla,McGough:2013gka}), the quantum dimension of a non-degenerate operator with weight $\Delta_p=p^2+\frac{1}{4}Q^2$ reads, $d_p=\sinh\left(\pi p b\right)\sinh\left(\pi p b^{-1}\right)$. Moreover, the S-matrix element between two such non-degenerate operators is given by 
${\cal S}_p^{\, q}=\sqrt{2}\cos\left(\pi p q\right)$. Plugging these into \eqref{OTO late} yields at large-c
\be
C_{pq}^{\,\beta}(t)\sim \Lambda_{p,q}\,c\, e^{-\pi\left(p+q\right)\sqrt{\frac{c}{6}}}\,,\nonumber
\ee
where $ \Lambda_{p,q}$ is a constant that depends on $p$ and $q$. Observe that the above expression is damped exponentially as we increase the central charge in contrast with the RCFT case.

\end{document}